# Starry Messages: Searching for Signatures of Interstellar Archaeology

Richard A. Carrigan, Jr.,
Fermi National Accelerator Laboratory*, Batavia, IL 60510, USA
carrigan@fnal.gov
630-840-8755

Summary

Searching for signatures of cosmic-scale archaeological artifacts such as Dyson spheres or Kardashev civilizations is an interesting alternative to conventional SETI. Uncovering such an artifact does not require the intentional transmission of a signal on the part of the original civilization. This type of search is called *interstellar archaeology* or sometimes *cosmic archaeology*. The detection of intelligence elsewhere in the Universe with interstellar archaeology or SETI would have broad implications for science. For example, the constraints of the anthropic principle would have to be loosened if a different type of intelligence was discovered elsewhere. A variety of interstellar archaeology signatures are discussed including non-natural planetary atmospheric constituents, stellar doping with isotopes of nuclear wastes, Dyson spheres, as well as signatures of stellar and galactic-scale engineering. The concept of a Fermi bubble due to interstellar migration is introduced in the discussion of galactic signatures. These potential interstellar archaeological signatures are classified using the Kardashev scale. A modified Drake equation is used to evaluate the relative challenges of finding various sources. With few exceptions interstellar archaeological signatures are clouded and beyond current technological capabilities. However SETI for so-called cultural transmissions and planetary atmosphere signatures are within reach.

ACKNOWLEDGEMENT
*Operated by the Fermi Research Alliance, LLC under Contract No. DE-AC02-07CH11359 with the United States Department of Energy.

# 1. INTRODUCTION

In *Starry Messenger* [1], Galileo tells how he used one of the first telescopes to discover Jupiter's moons. Now, four centuries later, these objects have been observed in great detail by planetary probes. Almost as exciting, extra-solar planetary atmospheres have been detected [2]. There are hopes for observing atmospheric signatures of life. Mission planners for future exoplanet atmosphere studies even ask if signs of intelligence could be detected [3].

The search for exoplanet atmospheric markers of intelligence is one example of *interstellar archaeology* or *cosmic archaeology*. Another example is the search for a Dyson sphere [4]. The Dyson sphere conjecture speculates that a planet could be purposely broken up to form a heat absorbing shield around a star to provide more useful energy. One fanciful model of a Dyson sphere would be a star enclosed in a shroud of solar-cell calculator chips. A recent whole sky search for Dyson spheres using the database from the IRAS satellite [5] has shown that there are at most only a few lackluster candidates in a region containing a million suns. A search for Dyson spheres is an illustration of a directed search for intelligence assuming a specific hypothesis. An alternative is to search for extraterrestrial intelligence (SETI) by looking for actively generated signals such as radio waves. A search is now underway at the Allen Telescope Array to look for radio signals from a sample of a thousand older stars that are plausible candidates for harboring life along the lines here on Earth.

These are three examples of interstellar archaeology. Other possibilities include searching for non-natural components in the atmospheres of stars or looking for examples of stellar or astro-engineering. Some of these possibilities have been discussed in Lemarchand's seminal work [6] on the detectability of technological activities beyond the earth.

SETI and interstellar archaeology are parts of a wave of developments that have emerged in the last decade or so under the umbrella of the search for life in the universe. Even particle physics and cosmology are linked to this search through the anthropic principle [7], the idea that only a universe with physical laws tuned to the requirements of life as it exists on Earth would be suitable for mankind. Other developments have included the enormously broadened picture of the solar system, a deepening understanding of the emergence of life on Earth and the ability of so-called extremophiles to adapt to hostile environments perhaps including such unlikely places as Mars [8], Titan [9] and Enceladus [10] as well as the discovery and investigation of large numbers of exoplanets.

The paradigm that underlies interstellar archaeology, indeed archaeology in general, is different than the one that ordinarily drives radio and optical SETI. Much of the SETI program is predicated on the originating civilization making a decision to transmit a radio or optical signal. This might be a beacon that mainly says "look at me!" In the case of archaeology an arrow head, a pyramid, a sunken Titanic, or a Dyson sphere was created for a purpose and only incidentally became a peephole to look back in time to another civilization. Clearly SETI may stumble on a interstellar TV program advertising cigarettes or astronauts could find a Rosetta stone on Mars. But generally, signals suitable for SETI require a motivated sender while archaeological artifacts were motivated by other circumstances such as hunting or building monuments to gods.

While interstellar archaeology signatures will generally not have been motivated by the urge to communicate, they may have been engendered by other considerations. Interstellar archaeology signatures could be driven by habitable zone change [11], stellar catastrophes such as a nearby supernova, going off the main sequence to become a red giant, or aging planets with evolving bad atmospheres. As a star proceeds through its evolutionary path it eventually drops off the main sequence and perhaps on to the asymptotic giant branch (AGB). If it turns into a red giant the stellar atmosphere may reach out and engulf the planets. Are there "cosmic engineering" strategies that an advanced civilization could take to escape this disaster such as migration to a nearby star? Would these efforts lead to interstellar archaeological signatures? This topic has been discussed by Beech [12] in a book where he gives illustrations of some possible engineering solutions to extending the life of a star. Beech [13] has recently noted that planet-star systems near the end of their main sequence lifetime are logical places to look for evidence of planetary engineering. Zuckerman [14] estimates up to 10,000 civilizations in our galaxy may have already faced this challenge.

Even now there are some primitive examples of archaeology across the void of space. Extraterrestrials (astronauts) in the International Space Station routinely observe pyramids on Earth that are more than four millennia old. The nearly forty year old Apollo 15 landing site on the moon has been photographed from space. The Mars Global Surveyor has been able to photograph the Mars Spirit Rover landing track, as well as the rover itself and the parachute and heat shield. Since 2005 Venus Express, now circling Venus, has been used to look for signs of life on Earth. Earth only occupies one pixel in the imaging sensor so that the task is harder than it might sound. In fact, researchers such as Carlotto [15], Sivier [16] and Dietrich and Perron [17] have already given some thought to distinguishing man-made artifacts and other signs of life from natural phenomena both on Earth and Mars.

This article focuses on larger scale projects. Searches for artifacts within our solar system such as artificial Kuiper belt objects [18] have been suggested. These are not discussed.

## 2. THE KARDASHEV CLASSIFICATION SYSTEM

Almost at the beginning of modern interest in SETI Kardashev [19] proposed a classification scale for advanced civilizations based on their use of energy. Kardashev appropriately used Roman numerals which follows the convention of using them as an index. Type I civilizations might harness all the stellar power falling on the surface of a planet, $2*10^{17}$ W for Earth. (Kardashev used $4*10^{12}$ W, the "level presently attained on the Earth." World power consumption is $\sim 2*10^{13}$ W [20].) A Type II civilization would consume the power of a star ($4*10^{26}$ W for the Sun). A full Dyson sphere is an example of a Type II civilization. A Type III civilization would exploit the power of a galaxy (on the order of $4*10^{37}$ W). Fig. 1 is a schematic illustration of the Kardashev scale.

Several amplifications to the Kardashev scale have been suggested. To accommodate cases where a civilization had not yet fully exploited available energy resources Sagan [21] introduced a logarithmic interpolation:

$$K = \frac{\log_{10} W - 6}{10}$$

where K is a decimal Kardashev index and W is the power output for the civilization. Others such as Galanti [22] have suggested an extension of the scale to Type IV in order to include the visible universe. The volume defined by the comoving distance to the edge of the visible universe (14 billion parsecs) is about $10^{31}$ cubic parsecs. It contains on the order of $10^{13}$ large galaxies. This could be extended even further to Type V, to include the concept of multiverses (see later section). By the same token one could ask about smaller civilizations, for example a town of 10,000 people (Type 0; 100 W/person), several thousand fleas (Type -I; $10^{-4}$ W), or mycoplasma genitalium (Type -II; $10^{-14}$ W). An individual mycoplasma genitalium contains on the order of a billion protons. Note that the Romans had no form for negative numbers, decimal fractions, or zero (a version was introduced later). For the Sagan scale an appropriate notation might be a Roman integer with a minus sign if needed followed by an Arabic decimal fraction. Interestingly going from Type –II (sub-micron microbes) to Type IV (the visible universe) spans a range of $10^{60}$ in mass and power if mass is a rough proxy for power. This bears some relationship to the famous observation that there are $10^{80}$ protons in the visible universe.

Earth itself is an illustration that civilizations can approach a Type I Kardashev classification. Amazingly, the Earth is only four orders of magnitude away from a power use that would qualify as Type I. Other examples of near Type I signatures are non-natural constituents in extrasolar planetary atmospheres and evidence for nuclear fuel disposal in a star. For example a prominent $CO_2$ line in a planetary atmosphere might suggest heavy use of fossil fuels. In the case of Earth, operating at about a ten thousandth of a Type I civilization, the $CO_2$ fraction in the atmosphere has increased by 35% in the course of the Industrial Revolution probably due to heavy fuel use. Many other activities that could be extraterrestrial signatures of civilization involve power outputs well below Type I. These include radio and TV transmissions, visible night lighting, rocket launch trails and the explosion of nuclear weapons.

The recent Dyson sphere search [5] using IRAS may well have been the first systematic whole-sky search for a Type II signature. Dyson sphere searches are discussed later in the article. About a decade ago Annis [23] at Fermilab looked for Type III civilizations by searching for dark galaxies. This is discussed later in the section on galactic-scale signatures.

## 3. THE DRAKE EQUATION

The foundation for a scientific approach to SETI is the Drake equation. Drake devised the equation to organize the results from perhaps the first meeting on SETI. The equation appears in several forms. Typically it is used to calculate N, the number of communicative civilizations in the Milky Way galaxy. One version is

$$N = Rf_p n_e f_l f_i f_c L_c$$

where R is the formation rate of intelligent life-friendly stars, $f_p$ is the fraction of these stars with planets, $n_e$ is the average number of planets in a planetary system that are hospitable to life, $f_l$ is the fraction of these planets where life emerges, $f_i$ is the fraction of these planets on which intelligent life arises, $f_c$ is the fraction of these where an interstellar-worthy civilization emerges and $L_c$ is the length of time the civilization remains detectable.

From the outset this equation has been seen as a guide to the factors going in to SETI rather than a precise statement of the probability of detecting signs of extrasolar intelligence. Some of the factors may be mutually linked so that they are not independent variables. As an illustration, Livio [24] notes that the timescale of biological evolution and the lifetime of the host star are not necessarily independent. Some other factors may be missing. For example, there is not an average stellar lifetime in the equation. The use of the word "civilization" may subtly understate possibilities related to robotics and computers. Nevertheless, the Drake equation is a useful yardstick for understanding SETI and where the interesting questions are.

A second set of factors is not embodied directly in the Drake equation. These concern the "degree of difficulty" for a civilization in executing something that may relate to factors in the Drake equation. For example, building a Dyson sphere out of the shards of a planet is a challenging project. Engineering the useful lifetime of a star is even more so. Possible proxies for the degree of difficulty are the energy harnessed or the mass involved.

The variable that is most directly related to interstellar archaeology is $L_c$, the time the civilization is detectable. To emphasize this point the Drake equation for some interstellar signature, x, can be rewritten as:

$$N_x = f\frac{L_x}{L_c}$$

where $L_x$ is the lifetime and $f = Rf_p n_e f_l f_i f_c$. For some cases of interstellar archaeology this lifetime, $L_x$, could be substantially longer than the lifetime of the originating civilization. To roughly paraphrase Gautier's famous lines "art alone endures: the bust outlasts the throne, the coin, Tiberius." Some of these interstellar archaeology cases are discussed below.

## 4. **SETI**

SETI and the search for interstellar archaeology are not competing programs but complimentary and related efforts. The most common paradigm for SETI is to look for an intentional signal from a radio or optical beacon. Typically this is assumed to be a narrow-band electromagnetic carrier in an obvious place in the radio or optical spectrum. Such a signal with little or no signal content can have the maximum range. It is noteworthy that signals from relatively nearby stars might be less constrained by the need for powerful narrow band, narrow cone antenna-transmitter systems and therefore could be more information-rich. Nearby transmitting civilizations might also be in a better position to beam signals to a limited number of suitable candidate stars or even have gleaned evidence of intelligent activity on Earth from their own interstellar archaeology searches directed at our solar system.

Our civilization now emits background radio emissions. Some of the transmissions from Earth are beamed while others are more or less omni-directional. A civilization fifty light years away detecting Earth's first radio signals could have broadcast a return signal that would now be reaching Earth. There are about 400 stars within this fifty light year sphere. Tarter [25] has estimated that an ordinary strong TV transmission from Earth could be detected by our current technology out to about a light year. With the Square Kilometer Array (SKA) the sensitivity would increase more than 10 fold so that the reach could be 4 light years, a volume encompassing a star or two.

A radio SETI signal search can be narrowed by using information about habitable zones, stellar age and increasingly even information concerning exoplanets around a star [26]. Indeed results from interstellar archaeology investigations can also help to delineate SETI search strategies. In this spirit an investigation of 13 of the least implausible Dyson sphere candidates in the recent IRAS search [5] is planned for the Allen Telescope.

Searching for a radio or optical SETI message, directed or omni-directional, intentional or cultural, is a form of archaeological investigation. Ordinary radio and optical searches expect to have their greatest sensitivity for directed beams with narrow bandwidth signals. Directional transmission from the Arecibo planetary radar could be detected out to 3000 ly. Note that there is a problem with looking for leakage radiation since the signal strength is low compared to a directed source. In addition as a civilization progresses the emission of stray radiation may be much more limited. Some years ago Sullivan and Knowles [27] investigated this challenge of looking for leakage radiation by detecting radio signals bounced off the moon. In any case, current SETI searches will continue to probe the radio and optical sky and may at some point detect cultural as distinguished from intentional signals.

Over the last decades several wide-ranging searches have been made for radio ETI signals [25]. No interesting signatures have been found that could be replicated. The searches using the Arecibo 300 m diameter radio telescope are among the most interesting. The Phoenix search has targeted about 1000 stars within 250 light years principally selected for characteristics that might be hospitable for life**.** The system monitored the band from 1.2 to 3 GHz. This band includes the "water hole" extending from the neutral hydrogen line at 1420 MHz to the OH radical line at 1720 MHz. A region from -2º to +38º declination in celestial coordinates was searched with a sensitivity of 1 Jy. No ETI signals have been observed in the Phoenix survey. At Arecibo the Berkeley group that operates the SETI@home program has carried out a commensal survey with other Arecibo programs. This program, SERENDIP [28] has collected data in a side lobe of the main Arecibo beam. As a result they had full coverage over the Arecibo portion of the sky but the sensitivity has been an order of magnitude lower than Phoenix. Like Phoenix they have detected no ETI signal.

Motivation is an interesting question for intentional SETI signals. This is not a consideration for most other examples of interstellar archaeology. Specifically, a message from an extrasolar civilization could have an agenda behind it. This agenda might not necessarily be positive. Indeed, it might be malevolent. A somewhat parallel case, biological contamination from space samples, is a remote but accepted possibility. Could a similar problem occur with information from an extrasolar civilization? Many consider this "SETI Hacker" conjecture [29] deeply flawed [30].

The challenges of both originating and receiving SETI signals is modest compared with many other examples of interstellar archaeology. In his thesis Leigh [31] estimated that sending a word 4000 light years over a directed link using 200 m dishes might cost about $1 at $0.1/KWh. Sustaining the activity may be more of a problem. There are relatively few illustrations on earth where a technology-based signal has continued for more than a century or so. In any case when a civilization dies any transmission will probably cease. A second problem concerns the duty factor for transmission. For example one transmitter might be used to beam signals at a number of different stars. The transmitter might only be "visible" for a portion of the extrasolar planetary day. Some of these problems could be solved by a diligent transmitting civilization.

In summary, the current reach of radio SETI programs is in the range of 250 to 1000 light years. The advent of the Allen telescope will effectively result in much more telescope time devoted to SETI work. That will increase the number of surveyed sources by a factor of several hundred and the range to the center of the galaxy and beyond.

Some of the relevant factors for SETI searches are summarized in the Table of various examples of interstellar archaeology. The table includes rough estimates for the reach, associated lifetimes, power requirements and mass involved as surrogates for the degree of difficulty of the search.

## 5. EXOPLANET ATMOSPHERES

Could synthetic or unnatural constituents in an exoplanet atmosphere serve as an interstellar archaeological marker? Freon®, part of a family of chloro-fluorocarbons, is an example of such a contaminant. The recent observations of exoplanet atmospheres have now opened the extraordinary possibility of searching for non-natural signatures.

More than 400 exoplanets have now been discovered. The limits of current observational techniques introduce biases into the distributions for the masses and periods of planets that have been found. Even with the challenges of instrumental biases there have still been major surprises such as the existence of so-called hot Jupiters. In the next decade several space missions should expand the exoplanet sample considerably. Both the French-ESU COROT satellite launched in 2007 and NASA's Kepler are now in space. COROT and Kepler look for terrestrial-scale planets by detecting small reductions in the stellar light when a planet transits in front of the star. These missions should easily double the number of known exoplanets.

Exoplanetary atmospheres have been detected by watching the spectral behavior of a planet-star system as the planet transits the star. Employing this technique Charbonneau [32] and his colleagues used the Space Telescope Imaging Spectrograph (STIS) on the Hubble Telescope to make the first detection of a sodium atmosphere around Osiris b (HD 209458b). Another group [33] using the STIS to look at Osiris saw spectral lines from H, C and O absorbed by the atmosphere as well as evidence for a tail behind the planet due to atmospheric loss. Richardson et al. [34] using the Spitzer infrared spectrograph reported only the presence of a 9.65 micron peak from the atmosphere suggesting silicates and the possibility of a molecule with similarities to benzene. Missing from any of these measurements on Osiris so far has been any sign of water vapor or methane [35]. A very interesting development has been the detection of methane, water and carbon dioxide [36] in the atmosphere of another exoplanet, HD 189733b some 63 light years away. With current technology there is an approximately 10 $\sigma$ signature of $CO_2$ at 2.1 $\mu$m in the HD 189733b atmosphere. A 2 $\sigma$ signature might be seen out to 140 light years. Clearly these techniques for investigating extrasolar planetary atmospheres have turned out to be surprisingly powerful. They can already be used to eliminate atmospheric models.

Burrows [37] and others have developed models for “natural” extrasolar planet atmospheres. Detailed knowledge of planetary atmospheres in the Solar System is useful for extrasolar planetary atmosphere modeling. As a cross check recent observations of the visible and near infrared spectra of the moon during an eclipse have been used to determine the transmission and reflection spectral features of the Earth’s atmosphere in a way that simulates exoplanet transits [38].

An important step along the way toward interstellar archaeology is finding signatures for life in an exoplanet atmosphere [39]. About a decade ago Perryman [11] examined atmospheric signatures for life that might be found by imaging planets and concluded it is a difficult challenge. Carbon-based life would seem to require water in a stable environment. $O_2$, $O_3$ (ozone) and $CH_4$ (methane) have been investigated as atmospheric signatures of life. In this vein Lovelock [40] suggests that all the molecular oxygen and ozone gas content in Earth’s atmosphere is due to biogenic processes.

As noted earlier, the nature of an extrasolar planetary atmosphere might give some clue to the existence of intelligence. What is needed is a unique signal such as a Freon line that indicates intelligent activity. However, positive identification of a Freon molecule with low resolution spectroscopy could be challenging [41]. Better understanding of the composition of Earth’s atmosphere and the changes with time might help to identify extrasolar system signatures of the existence of a civilization. For example the carbon dioxide fraction in Earth’s atmosphere due to heating and the use of internal combustion engines could signal the activities of civilization on the earth. The $CO_2$ fraction by volume has risen from 284 ppmv in 1832 to 384 ppmv in 2007. This amounts to a $CO_2$ change of $8*10^{14}$ kg or $1.3*10^{-10}$ of the mass of the earth. Some suggest that at least part of the atmospheric $CO_2$ rise could come from natural processes and non-intelligent life. This is illustrative of the ambiguity of putative signals of interstellar archaeology.

In summary, while remarkable exoplanet atmospheric signatures have been obtained for several cases there is still some distance to go before there is enough sensitivity to find meaningful signatures of life and intelligence. COROT and Kepler should double the number of identified exoplanets in the next three years. However even the next rounds of satellites and observatories will probably not reach the necessary spectral sensitivity to clearly identify signatures of life or intelligence. On the other hand progress is also being made in modeling atmospheres and using lunar eclipses to infer the spectral observation conditions for exoplanets.

The appropriate lifetime for the equivalent Drake equation for an exoplanet atmosphere signature would be the life of the detectable atmospheric perturbation introduced into the atmosphere. This could be some combination of the time to inject the contaminant into the atmosphere and the time for the contaminant to disappear from the atmosphere.

## 6. STELLAR SPECTRAL SIGNATURES

The challenge of looking for optical spectral signatures from planetary atmospheres is that the light output is miniscule compared to the signal from the host star. An alternative interstellar archaeology strategy might be to find signatures from the host star that reflect the presence of intelligence. A signature could result from a process such as waste disposal in the star or even active salting of the stellar envelope with artificially produced isotopes. Processing by life itself can change an isotopic ratio. This happens with the ratio of carbon 12 to carbon 14 used in radioactive dating of organic material. Another intriguing possibility is modulating a stellar maser or a stellar optical signature with something such as clouds.

### 6.1 Stellar salting

The idea of actively salting a star was originally independently by Drake [42] and Shklovskii [43] in the context of possible active signaling. Independently they suggested that feeding a short-lived nuclear species with a strong resonant absorption line into a stellar atmosphere might be used to signal the presence of an advanced civilization. A search for such a signature would be in the spirit of looking for a radio or optical signal transmission rather than searching for astroengineering. For radioactive salting the signature has to arise from an atom or atomic isotope not normally present in the star. Drake suggested technetium (Z = 43) for this purpose. Technetium is the element with the lowest atomic number with no stable isotope. The other possible unstable atom in the low to medium Z range is promethium with Z = 61. The half lives of promethium's isotopes are in the year range and may be too short for signaling over an extended period of time.

Technetium is found in trace amounts in red giants and also on Earth where it is produced by uranium fission. Typically the red giants are relatively cold, variable S stars in the so-called third dredge-up period (3DUP) after hydrogen burning. There are a number of isotopes of technetium with lifetimes ranging from shorter than an hour to millions of years. A technetium line would probably be used as a beacon. To have a higher data transmission rate one would pick the shortest lived isotope but this would require a larger quantity of the dopant.

The reach for observation of a technetium signal is quite good. Recent optical observations by Uttenthaler, et al. [44] find stars containing technetium spectral lines in the galactic bulge 8.5 kpc away. Clearly this range depends on the concentration of the contaminant. Indeed the extraction of the technetium lines in the Uttenthaler paper involves a very significant analysis employing challenging observations with the ESO Very Large Telescope. Note that the properties of stellar atomic technetium lines are also difficult to calculate [45].

Lemarchand [6] using an approach employed by Drake estimated that about 100,000 tons of technetium would be required to produce a recognizable spectral signal. For roughly the same conditions Drake estimated 4000 tons. This is a large amount of technetium. Even so this estimate of the amount may be rather low because it assumes each atom emits light at the maximum rate and absorbs like a black disc. Several examples give a sense of scale for putting $10^5$ tons of exotic material into a star. A plutonium bomb might weigh on the order of 10 kilograms. One solar panel carried aloft by the Space Shuttle weighed 18 tons. Reactors have produced on the order of 100 tons of technetium over the last 6-7 decades.

This approach seems like an unlikely candidate for an interstellar archaeology search because of the existence of trace natural technetium signals in some stars, the mass required and the fact that modulating the signal with an isotope with a shorter lifetime would require much more material.

## 6.2 Nuclear waste

A second search strategy suggested by Whitmire and Wright [46] is to look for signatures of nuclear fission waste products that have been disposed of in a star. Looking for such signals is the interstellar archaeological equivalent of searching ancient waste dumps for cultural artifacts. Whitmire and Wright note that convective mixing and stellar lifetimes restrict the possible candidates for observation of nuclear waste disposal to stars in the spectral classes A5 to F2. Their point is that slow neutron fission products such as neodymium and praseodymium can result in anomalous abundances relative to normal stellar nucleosynthesis. Fission-produced unstable elements such as technetium and plutonium could also be markers.

Generating an observable signature could require the nuclear waste products derived from a substantial fraction of a planet's fissile material. In turn that material would have to be transported to the host star. As with isotope salting, there would be a high cost premium to move radioactive waste from a planet to a star, although the energy cost to move it would be small compared to the power yield that could be obtained from the material. Finally, it would not be possible to easily modulate the signal. As a consequence a line from waste would only serve as a cultural marker. Identification of a candidate would rest on abnormal line intensities, not the presence or absence of a spectral line.

In most essentials concerning scale, the Drake equation for the nuclear waste case follows isotope salting.

## 6.3 Isotopic ratios

Isotopic ratio such as $^{26}Al/^{27}Al$ and $^{12}C/\,^{14}C$ are used as dating tools for the solar system and archaeology on the Earth. Neither of these ratios is germane as a sign of intelligence near a star. Still, there are other possible isotopic anomalies such as those from nuclear waste disposal discussed above. There is also a possibility that some form of stellar engineering might change isotopic ratios.

Some atomic isotopic spectroscopy splittings are in the range 0.1 - 0.2 Angstroms (0.01-0.02 nm). This can be resolved with a high resolution spectrograph on a telescope [47]. Once again the problem is that the signature is modulated by natural effects. For example, Cowley et al. [48] find that the isotopes of heavy metals will be selectively segregated in chemically peculiar (CP) stars with the heavier isotopes rising in the atmosphere because of isotope dependent light-induced drift, somewhat in the spirit of laser isotope separation. The root causes of these effects are often complicated involving such factors as the stellar atmosphere profile and magnetic fields. As a result it would probably be quite difficult to make a case for an artifact due to a non-natural effect involved with stellar engineering.

## 6.4 Spectral modulation

Another intriguing possibility is to look for evidence of intelligent modulation of cosmic lasers and the optical and infrared signatures of carbon stars. In the relatively late phase of stellar evolution carbon stars are surrounded by dust clouds that modulate both visible and infrared signals. Typically the mass loss rate is $10^{-6}$ stellar masses per year. Of course the environment of a carbon star is probably not conducive to life [49]. As noted earlier, in the red giant phase Earth will most likely be swallowed as the sun expands. On the other hand if there was some possibility of life continuing the situation might engender a spirit of grand engineering and also an urge to communicate. In many cases the red giant environment generates varying maser signals. Modulation could emerge from dust clouds [50] moving and transforming in the spirit of weather systems on the Earth. ("Dust clouds" here is used to describe dust clumps around a star.) Modulation could also arise from linking the magnetic field from a Jupiter-scale planet and the stellar equivalent of the solar wind [51]. A "movie" produced using a very long base line interferometer [52] shows a spectacular demonstration of natural modulation of a number of maser spots around the Mira Variable TX Cam.

The solar wind produces magnetic anomalies of order 1% on Earth in conjunction with the aurora. High altitude nuclear explosions such as the Argus tests of the late fifties and early sixties can produce geomagnetic effects and artificial Van Allen belts that last several weeks. Something along this line might be used to produce temporary planetary magnetic field variations. Perhaps visible and infrared modulation could be produced by seeding carbon dust clouds or by trying to raise and lower them by inducing surface heat transfer over a small part of the stellar area.

A limitation here is to find a unique trait that is a signature of intelligence. Red giants already exhibit a variety of different modulations. The overall process has the flavor of the child's game of looking for faces in clouds. One handle might be to look for modulations on very short time scales that would not occur with ordinary physical processes.

To summarize, the possibility of observing a stellar spectral effect from some non-natural cause will be severely hampered by the diversity of natural effects. In many cases, these natural effects will be difficult to model and observe in their own right. By current standards these techniques require large quantities of rare isotopes or large scale engineering.

The equivalent Drake equation for stellar spectral signals varies from case to case. For some cases involving salting, nuclear waste and isotopic ratio modification the marker could persists for very long times so that for that case

$$N_{sp} = fL_{sp} / L_c$$

where $L_{sp}$ is the lifetime of the spectral signature. On the other hand modulation of stellar masers or optical spectral lines would be more like conventional SETI. That is to say, the lifetime of the modulation signature could be the effective life of the transmitting civilization. For cases that occurred during a red giant phase this lifetime would be the lifetime of the red giant phase, often a short time. An important point for all these cases is that a different form of the Drake equation holds for each of them where one factor has been replaced by another.

## 7. **DYSON SPHERES**

Dyson spheres [4] are among the more intriguing possibilities for interstellar archaeology signatures. The swarm of tiny objects in a Dyson shell would greatly increase the useful area for advanced activities and absorb all of the visible light. The stellar energy would be reradiated at lower temperature. If the visible light was totally absorbed a pure Dyson sphere signature would be an infrared object with luminosity equivalent to the invisible star and a blackbody-like thermal distribution with a temperature corresponding to the radius of the shell. This is similar to a carbon star surrounded by a thick, low temperature dust cloud. The assumption of a fully-shrouded body is restrictive. A partial Dyson sphere, for example a ring, would be a more practical object to build. By looking for infra-red excesses around visible stars several astronomers including D. Werthimer [53] and J. Jugaku [54] have searched for partial Dyson spheres where the star is only partly obscured. No likely candidates were reported in searches of several thousand stars.

The mass of a Jupiter-scale Dyson sphere would be $2*10^{27}$ kg. Earth's biomass is in the $10^{15}$ kg range. The largest ocean liner is $1.5*10^{8}$ kg while the immobile Great Wall of China is on the order of $10^{12}$ kg and took 2500 years to build. The International Space Station weighs $2*10^{5}$ kg so that it is a factor of ten billion trillion smaller than a Jupiter-class Dyson sphere. In short, a Dyson sphere is a large object! Dyson's perspective on this is to note that increases of scale such as this can be reached in 3000 to 4000 years based on a population growth of O(1%/year).

The energy cost of constructing a Dyson sphere is high. In his paper Dyson notes that the energy to disassemble and rearrange a Jupiter size planet is equivalent to 800 years of <u>total</u> solar radiation, not the much smaller amount of solar radiation falling on the planet's surface. (The 800 year number is of the order of Jupiter's gravitational self energy.) This poses a bootstrap problem in the building process. Initially the available power would be low but could rise as more surface area developed. In any case the construction time would probably be much longer than $10^{3}$ years.

One of the criticisms of Dyson's proposal has been based on the incorrect presumption that Dyson's object was a rigid, hollow sphere. Such a large object would be unstable. This argument ignores the possibility of a shroud consisting principally of thin silicon chips, solar panels, or carbon nanotubes. Put differently, it is difficult to anticipate the size or the technology going in to the individual cells in a Dyson shroud developed by an advanced intelligence. Another criticism of the Dyson hypothesis is that a spherical swarm is not a stable system. On the other hand the individual cells in the shroud could be actively steered, perhaps by solar sails.

At first sight a pure Dyson sphere is a clean signature for an interstellar archaeology search. However there are a number of natural astronomical objects which bear some resemblance to a Dyson sphere. To definitely identify a Dyson sphere one has to rule out the more conventional and plausible natural sources that are mimics. The birth and death phases of many stars are characteristically associated with heavy dust clouds around the star. Since the associated lifetimes of the processes giving rise to these clouds are short, the objects are rare. Many of these sources show silicate emission while others exhibit silicate absorption. They may behave as natural masers or show pulsations.

The IRAS spacecraft that flew in the mid-eighties was almost an ideal instrument for a Dyson sphere search. It covered nearly the entire sky. Importantly, IRAS carried a low resolution infrared spectrometer (LRS) in addition to a set of four filters. The spectrometer sensitivity was 1 Jansky while the angular resolution was 1 minute. Two hundred fifty thousand point sources were identified. A comprehensive atlas of more than eleven thousand sources containing a significant portion of the available useful IRAS Low Resolution Spectra has been compiled by a Calgary group [55].

Recently Carrigan [5] has used information from the IRAS LRS database to build on earlier IRAS filter-based searches by several astronomers including Kardashev [56] and Slysh [57]. The IRAS-LRS search used a sequential series of cuts to select a set of objects that could be Dyson sphere candidates. The cuts included factors such as temperature and a scan of the spectral distribution to look for features including spectral lines or a blackbody-like spectrum. Fig. 2 shows a common signature for a low temperature carbon star with a silicon carbide line. The line rules this out as a Dyson sphere. For this source there is a distinct difference between the filters and the LRS, likely due to source variability. This is a source that is barely inside the 600 ºK upper temperature bound used in the search.

A source with a well-established classification was generally discarded unless there was some reason to doubt the classification. The cuts along with statistical tests narrowed the sample to sixteen mildly interesting sources. Only three of these had relatively low spectral statistical fluctuations. All of the sixteen sources have some feature which clouds their identification as a Dyson sphere. In practice, most of the LRS candidates have higher temperatures and just don't look much like the spectrum expected from a Dyson sphere. The search suggests that there are few if any even mildly interesting candidates within several hundred light years of Earth.

Fig. 3 is the spectral distribution for one of the more interesting examples, IRAS 20369+5131. This source is one of the best fits to a Planck distribution. The temperature is around the boiling point of water and there is no visible star. The SIMBAD catalog classifies this as a carbon star while Calgary classifies it as U, an unusual candidate.

The largest one-Sun bolometric distance in the 16 source sample is 118 pc. This was for a maximum peak flux of 8.6 Jy. (The one Sun bolometric distance is defined in [5] based on an equation developed by Slysh [57].) The 11,000 source Calgary sample extended down to 1-2 Jy which would have given a maximum one-Sun bolometric distance of 300 pc for LRS sources. This region includes something in the neighborhood of a million stars. For a galactic scale height of 100-200 pc the bolometric distances of the most interesting sources suggests they should be scattered over the whole sky rather than lying near the galactic plane. As seen in the galactic Aitoff plot in fig. 4 only one quarter of the 16 sources are scattered over the sky indicating that many of the sources are at greater distances and are more luminous. Fig. 4 also shows several thousand sources with the classifications and temperatures selected for the search. They follow the galaxy but show no bulge at the galactic center suggesting they are closer than the center.

In summary a Dyson sphere does not require intent to communicate on the part of a civilization. The current detection reach is comparable to a SETI search. However there is a problem of confounding signatures from mimics such as carbon stars. Searches for potential Dyson spheres would be sharpened by developing more realistic pictures of construction scenarios including such factors as time to build and approaches to stability. (Parenthetically M. Fogg has prepared a relevant bibliography [58] on terraforming and astroengineering.) Finally it would be interesting to consider how stellar evolution might stimulate the necessity of such large scale structures with a view to looking at candidate objects in the later stage of evolution along the main sequence.

The equivalent Drake equation for a Dyson sphere is

$$N_{Dy} = fL_{Dy} / L_c$$

where $L_{Dy}$ is the lifetime of a typical Dyson sphere. This could be similar to a deterioration time for an archaeological ruin or it might relate to the collapse of the sphere if active control was turned off. Characteristically $L_{Dy}$ could be substantially longer than the lifetime of a civilization.

## 8. STELLAR ENGINEERING

Since its birth 4.6 billion years ago Earth has led a relatively tranquil existence. While solar luminosity may have increased by 40% the Earth's surface temperature has remained essentially constant, probably due in part to the mitigating effect of the atmosphere [59]. The next five billion years will not be as hospitable. A main sequence star such as the Sun becomes cooler and redder as it burns its core hydrogen. The stellar surface expands and the luminosity increases. In *Rejuvenating the Sun and Avoiding Other Global Catastrophes* [12] Beech notes that roughly speaking a one percent change in luminosity will lead to a rise in temperature at the Earth's surface of one degree Kelvin. In five billion years the solar luminosity will increase by a factor of two to three, probably enough to destroy life. After that the Sun will enter a red giant phase in which the stellar radius expands by one hundred so the Earth is engulfed by the Sun. Characteristically by the red giant phase the star has only burned several percent of its hydrogen. The details of a star's path along the main sequence and into the red giant phase depend on factors such as the stellar mass and metallicity. Obviously the fate of the planets associated with a star also depend on all the parameters of the planetary system as well as the stellar history. (For a recent discussion of this see Villaver and Livio [49].)

At some time before the temperature rises too high our descendants will have to take drastic steps to avoid the fateful impact of rising luminosity and an expanding Sun. One possibility might be interplanetary or interstellar migration. Another potential option would be stellar engineering.

Are there astroengineering projects a civilization could undertake to engineer the path of the host star along the main sequence and thereby extend planetary life? A typical main sequence star turns into a red giant when the hydrogen of the inner core is exhausted. The useful stellar life could be extended by mixing the unused hydrogen in the outer envelope with the core. This is no easy astroengineering project. Beech [60] has studied the possibilities for this type of astroengineering in detail. His book [12] reviews some options. Beyond Beech's work there has been little recent technical discussion of astroengineering the Sun. Many, probably most astronomers view stellar engineering as an impossible task.

Beech examines a number of possibilities for engineering the stellar life cycle including mixing the stellar core with outer layers, inducing stellar mass loss, changing the pressure by adjusting the sun's rotation rate, or increasing the opacity by introducing heavy elements. He uses an equation for luminosity, L, which incorporates functional forms for the stellar temperature gradient and the opacity:

$$L = L_{KR} \frac{\mu^{7.5}}{(1+X)} M^5$$

where M is the mass of the star, μ is the mean molecular weight, X is the hydrogen mass fraction and $L_{KR}$ is a constant. Because the Kramer's expression for opacity has been introduced this formula for the luminosity differs from the more familiar version that goes as $M^3$. Beech's form is appropriate for stars in the Sun's mass range. Differentiating this to give fractional changes results in:

$$\frac{dL}{L} = 5\frac{dM}{M}.$$

This suggests that roughly 60% of the solar material would have to be dispersed to keep the luminosity constant to the end of the main sequence.

The tools Beech discusses for producing these changes such as mass loss are staggering. They include black hole pumps, ramjets and super-sized accelerators. In the framework of the twenty-first century none of these would be straightforward. For example, problems associated with induced mass loss would resemble some of the natural processes associated with the red giant phase.

For some sense of scale consider several absurd examples. The largest nuclear bomb exploded on Earth had an energy release of 50 megatons or $2*10^{17}$ J. The energy may have been released over $5*10^{-8}$ s so that the instantaneous power output was $4*10^{24}$ W or 1/100 of the power output of the sun. The outer pressure on the sun could be increased momentarily (opposite to the desired effect) by simultaneously exploding a very large number of bombs over the solar surface. But for a sustained effect this would have to be repeated ten million times a second. In the process visible light coming from the direction of the sun would increase many-fold. The point of the example is that very large inputs of power are required to affect solar changes. Another possibility is to drop the planet Mercury into the sun (again this would appear to be in the opposite sense needed to decrease the luminosity). That would bring some heavier elements such as sodium and oxygen into the solar surface and perhaps increase opacity starting at the surface. Leaving to the reader the details of how to drop Mercury into the sun, the net contribution to the solar mass would be on the order of one part in ten million or one hundred thousandth of the heavy element fraction. The point is that one of these engineering processes must proceed on a heroic and post twenty-first century scale if it is to happen at all.

In the discussions of the Dyson sphere the details of sphere assembly were put aside and instead one looked for signatures of the process. The same approach can be used for stellar engineering. For example, Beech suggested looking for hot, massive stars called blue stragglers that lie on an extension of the main sequence line beyond the AGB turn-off point to the red giant phase. He noted that some of these blue stragglers might be examples of astroengineering where the outer shell hydrogen had been mixed into the inner core to prolong the main sequence life. Blue stragglers had been something of a mystery although work using the Hubble telescope has shown that they could arise in “intimate” stellar encounters leading to coalescence of two stars in crowded globular clusters such as 47 Tucanae [61]. Recent population studies [62] lend additional weight to this picture. Beech recognizes the emerging consensus on a natural explanation for most blue stragglers but also suggests that one could look for isolated cases where the coalescence picture could not explain them.

The characteristic dense stellar globular cluster environment for blue stragglers also seems a less likely home for planetary life because of the disrupting effects of stellar collisions on planets. “Frequent” stellar collisions (say one every billion years) are bad for both planetary systems and life. Livio [63] notes that globular clusters are notoriously metal poor and are therefore not expected to be fertile hunting grounds for Earth-like planets. Livio goes on to ask “Why invoke astroengineering within globular clusters to explain what is a very natural blue-straggler phenomenon?” Now it probably makes sense to regard a blue straggler as a natural phenomenon that bears some resemblance to what a re-engineered star would look like.

In summary the grand scale of programs to homogenize the interior of a star or reduce luminosity by inducing stellar mass loss is overwhelming. This does not mean possibilities such as these should not be considered. For example, could the mass loss in the red giant phase be used as a mechanism to push a planetary civilization out to greater distances from the growing star? (Thompson, et al. contains a useful discussion of dust evolution in the red giant phase [64].)

Little is gained here by ungrounded speculation. Often a simple calculation will show that some putative project is untenable. Critical review by a serious astrophysicist is helpful. Still, the challenge of a SETI enthusiast or astrobiologist to the astrophysicist should be “What would you do, professor, when our raging sun is about to reach out and consume us?”

Seen from our twenty-first century perspective there are few, if any, astroengineering possibilities that are even mildly plausible. As a result there appear to be no very stimulating candidate signatures for astroengineering. Even with the enormous problems associated with extending the useful life of a star there is another way astrophysics can direct the search for astroengineering. If other extrasolar civilizations followed the behavior patterns of our cultures the most likely time for undertaking a heroic project would be when a situation was changing most rapidly. One way to search for intelligence –driven stellar activity would be to look at stars undergoing rapid evolutionary changes. In this picture a star close to the end of its main sequence lifetime might be a place to find signs of activities along the lines Beech suggests [13].

For very successful stellar engineering the Drake equation could take the form

$$N_{se} = fL_s / L_c$$

where $L_s$ is the order of the lifetime of a main sequence star. In some sense this would make an example of stellar engineering relatively more common than a SETI case. However this is one case where a new factor perhaps should be added relating to the degree of difficulty in carrying out the necessary engineering.

## 9. GALACTIC AND TRANS-GALACTIC SIGNATURES

Kardashev and others have proposed several pictures of possible Type III galactic-scale astroengineering projects. Typically these have been framed in terms of a Dyson sphere model. A galactic-scale Dyson sphere would require all the stars in the galaxy to build it leaving nothing left over as a heat source. This also ignores the cost of the energy to construct the object and other onerous problems. An alternative would be a galaxy completely populated with individual stellar Dyson spheres. Parenthetically, a Type III Dyson sphere object could act like a dark patch in the cosmos on a galactic scale with an associated infrared signature. This bears some relation to the topic of dark matter galaxies, a subject of interest in its own right.

### 9.1 Natural signals – dark galaxies

Dark matter was originally proposed seven decades ago by Zwicky [65] to explain the rotation behavior of galactic clusters. Zwicky found that the light yield in the Coma galaxy cluster was low compared to the mass expected on the basis of the virial theorem. Investigations of galactic rotation curves by Rubin [66] and others solidified confidence in the dark matter hypothesis. In the eighties dark matter became profoundly interwoven with particle cosmology. While there is general agreement that dark matter is very important and exists in nature there is still not an understanding of what it is. In recent years several galaxies have been suggested as possible pure dark matter galaxies, that is, galaxies containing only dark matter. An interesting but controversial candidate for a dark galaxy has been observed in Virgo about 60 million light years away (VirgoHI21) [67]. Critics have suggested that tidal shredding, not dark matter, explains the image. Several other relatively dark dwarf galaxies [68] such as the Ursa Major dwarf spheroidal UMa dSph [69] have been found recently. Any explanation involving ensembles of Dyson spheres would have to compete with other far more plausible possibilities. Many have suggested that dwarf dark matter galaxies should exist in large numbers but the data is inconclusive. Finally, examining candidates for dark matter galaxies requires detailed understanding of stellar populations and types as well as radio astronomy measurements to get neutral hydrogen spectral line information on either side of the galaxy and thereby determine the rotation rate. In this context, searching for an unnatural signature from galactic engineering is similar to looking for an eyeless needle in a disorganized haystack. The following discussion should be viewed in the light of this backdrop.

### 9.2 The Annis search

There has been at least one serious search for Kardashev Type III cases. Annis [23] has searched for Type III civilizations by looking for outliers on galactic distributions used to study the Tully-Fisher [70] or L-T scaling relation (for a recent summary see Tully's article in Scholarpedia [71]). The distribution of galaxies on a plot of galactic optical brightness or luminosity versus the maximum rotation velocity or radius of the galaxy follows a fairly consistent pattern. Cases lying below the typical galactic trend line reflect visible light that has been absorbed and emitted somewhere else in the electromagnetic spectrum. Annis examined existing distributions for spiral and elliptic galaxies and looked for sources below the normal trend lines where more than 75% of the visible light would have been absorbed. No candidates were found in a sample of 137 galaxies. From this Annis inferred a very low probability of a Type III civilization appearing that would be found using this search methodology. With more recent information [72] it is possible to extend the search to samples that are considerably larger and also more robust. It may be time to revisit this possibility. In particular with a sample ten to one hundred times larger one could examine the nature of a handful of outliers in more detail looking for any unique features such as a higher than normal infrared component. These outliers might be candidates for a radio SETI search.

### 9.3 Fermi bubbles

Dyson [73] notes that "a type III (Kardashev civilization) in our own galaxy would change the appearance of the sky so drastically that it could hardly have escaped our attention," while Annis [23] observes "It is quite clear that the Galaxy itself has not transformed into a type III civilization based on starlight, nor have M31 or M33, our two large neighbors." These statements are reasonable but what would happen for a civilization on its way to becoming a type III civilization, a type II.5 civilization so to say? If it was busily turning stars into Dyson spheres the civilization could create a "Fermi bubble" or void in the visible light from a patch of the galaxy with a corresponding upturn in the emission of infrared light. This bubble would grow following the lines of a suggestion attributed to Fermi [74] that patient space travelers moving at 1/1000 to 1/100 of the speed of light could span a galaxy in one to ten million years. Here "Fermi bubble" is used rather than "Fermi void", in part because the latter is also a term in solid state physics and also because such a region would only be a visible light void, not a matter void.

To get a sense of the time scale for a typical Fermi voyager recall that the escape velocity from the solar system from the vicinity of Earth is 42 km/s or ~0.0001 of the speed of light. It is notable that Voyager 1 is now traveling at a speed such that it will escape the gravitational potential of the sun. For an expenditure of 100 times the energy, the velocity could have been raised to 400 km/s. The diameter of the Milky Way is 60,000 light years so that a traveler going at a velocity of 400 km/s or $10^{-3}$ c would take 60 million light years to cross it. The sun takes roughly 200 million years to circle the center of the galaxy.

Newman and Sagan [75] modeled the possible diffusion of interstellar civilization across space. One observation is that the time to colonize individual stars is not large compared to the time to travel between stars. The expanding front of the civilization might move by a diffusive or random walk process. However it might also go directly to the next star so that the front would move forward at a rate comparable to the space travel velocity. In that spirit an intergalactic civilization could envelop a galaxy on a time scale comparable to or even somewhat shorter than the rotation period of the galaxy.

Searching for ETI-driven artificial bubbles in a spiral galaxy is challenging because the structure pattern itself is based on voids of a sort. The M51 Whirlpool galaxy seen face on at 30 mly illuminates the challenges. (A recent Spitzer study [76] is a useful guide to the M51 infrared distribution.) The sky area covered by M51 is about 6000 $kly^2$. A rough qualitative estimate suggests that there are no unexplained bubbles or voids at the level of 5% of the M51 galactic area. Transverse arm widths in the outer reaches are O(10 kly) so that it would be hard to identify void features below this level. Even if an interesting void in the visible with a corresponding infrared enhancement was identified for some galaxy it would be hard to make a case for a non-natural explanation. Interestingly Churchwell [77] and others have already established protocols and built catalogs to identify parsec-scale bubbles in the Milky Way produced by individual stars. Annis [78] has suggested that elliptical galaxies might be a better place to look for Fermi voids since they exhibit little structure.

## 9.4 Signatures transcending the galactic scale

Beyond the galactic scale lie galactic clusters, super-clusters and then penultimately, the visible universe. Roughly speaking the number of galaxies in the visible universe is the same as the number of stars in a galaxy. The overarching feature of the visible universe is the “Big Bang”, the putative genesis of the universe. Through most of history the universe has been viewed as the creation of a maker and, in some sense, the maker’s message to mankind. Now cosmologists grapple with the origin of the universe. A recent conjecture in natural philosophy is the combination of the so-called “anthropic principle” and a multiverse resting on string theory. The anthropic principle was first suggested by Brandon Carter [7]. (For a technical discussion see the article by Bjorken [79].) The idea is that the universe has to be just right for humans. It takes note of many curious coincidences in the physics including such processes as the origin of the elements. A particularly compelling example is the so-called triple alpha process discovered by Hoyle [80]. Multiverses in string theory open up the possibility of generating universes with an extremely wide range of different physical parameters. A characteristic string theory might offer up $10^{500}$ different models. In this picture mankind lives in a universe that yields the correct physical constants to give rise to the triple alpha reaction and everything else required for human existence. There is no message from a maker, only the happy but not altogether fortuitous circumstance that the universe is tuned just right for people.

## 9.5 Summary

In summary, there are several signatures for Kardashev Type III galactic –scale objects. By looking at Tully-Fisher and R-I-T scaling distributions Annis has set a limit on the fraction of galaxies that are dark that is below 1%. The Annis estimates could probably be improved by factors of ten using additional information that has accumulated. A problem here is that the signatures for so-called dark matter galaxies anticipated in many cosmological models bear some rudimentary relationship to a Kardashev type III signature. It also appears to be quite difficult to study the question of Fermi bubbles for spiral galaxies in any detail because natural bubbles and voids occur in galaxies. Elliptical galaxies might be a better place to start an investigation.

The lifetime, $L_c$, in the Drake equation for a Type III civilization would probably be at least on the order of the time to cross a galaxy. For relatively low energy outlays this could be tens of millions of years. R, the rate of formation of life-friendly stars, could have different values depending on the type of galaxy. Elliptical galaxies would have relatively older stars, increasing the opportunity for intelligent life to emerge but decreasing the formation rate of stars. For a galaxy saturated with Dyson spheres the ratio of the overall energy scale for a galaxy compared to a star would be just the number of stars in a galaxy, characteristically $10^{10}$. Note that this process would go on over a relatively longer time with conversion of one star being much like the next.

## 10. SUMMARY AND OBSERVATIONS

Year by year astronomy reaches further out into the Universe. Modern cosmology coupled with elementary particle physics brings human vision ever closer to a moment of creation. The understanding of astrophysical processes is becoming ever better so that even more is learned about the creation of the elements heavier than helium, the stuff of life and many planets. In the span of about a decade the understanding of the natural homeland of intelligent beings, planetary systems, has taken a giant step forward. The subject is approaching the interest level of cosmology or astrophysics. One can hope that soon signs of life may be detected beyond the Earth's solar system. Is it unreasonable then, to ask if searching for signs of intelligence in the Universe may lead to new insights?

Interestingly searching for intelligence beyond the solar system teaches people about themselves – how fragile the Earth's atmosphere and place in the universe are and how far the signs of civilization have already spread across the galaxy. The very detailed understanding of planetary atmospheres strikes a cautionary note about how Earth's global resources should be marshaled. Several of the most persuasive messages in this regard have come from Carl Sagan, one of the pioneers of modern studies of planetary atmospheres, astrobiology and SETI. SETI and interstellar archaeology help to set the measure of just what the so-called anthropic principle has to address. The constraints on the laws of a universe that the anthropic principle addresses would be loosened if life and intelligence in the Universe were found to be rich and diverse.

The search for interstellar archaeology highlights many of the difficulties and challenges facing the search for intelligence beyond the solar system. With several exceptions most of the examples of putative civilizations capable of executing projects on a grand interstellar scale require technology extrapolation so far beyond the present that it is difficult to imagine the situations at the source. The exceptions are active SETI and the study of extrasolar planetary atmospheres. Emerging SETI tools may be able to see cultural radio backgrounds many light years away. In the next several decades observatories should become available that will be able to study extrasolar planetary atmospheres in some detail.

Much of the discussion here has addressed signatures of interstellar archaeology. The emphasis has been on messages, not methods. Questions about how these signatures might come into being are important. For example, there has been little discussion of the details of building a Dyson sphere. In part this is because the projects are so fantastic. On the other hand after completing a backyard fence one can start to conceptualize building the Great Wall of China and still hope to gain perspective on the process. In another direction, Beech has outlined some of the possibilities for stellar engineering but much more could be investigated. In any case, it is clear that more work on processes leading to all the putative signatures would be helpful.

The Table ranks various examples of interstellar archaeology on the Kardashev scale. The values in the Table are intended only to be suggestive. The Kardashev scale does roughly correspond to the magnitude of the challenge or "degree of difficulty" faced by the corresponding civilization. Groups on Earth have already beamed purposeful radio messages to outer space, so-called active SETI. Planetary atmospheres are more challenging. Producing a detectable extrasolar signature requires a step up on non-natural atmospheric changes so far introduced in Earth's atmosphere by a factor of $10^3$. Stellar salting to cause optical spectral modification and nuclear waste dumping are well beyond any technology or resources available on Earth now. In addition the signatures would not shed much light on the circumstances of the originating civilization. Breaking up a planet to build a Dyson sphere completely transcends the will, ability and long term dedication of contemporary civilization. There is not even a mildly plausible technique for stellar engineering. Surprisingly the step from a Dyson sphere to a galaxy filled with Dyson spheres only involves a strengthened capability for space travel. On the time scale of Dyson Sphere construction this type of relatively slow space travel can be discussed even now. To recapitulate, SETI and the search for planetary signatures of life and intelligence currently appear to be the most accessible signatures. For Dyson spheres and stellar engineering work should mostly be directed toward investigating engineering techniques and limitations.

One measure of the likelihood of finding a signature of interstellar archaeology is the lifetime of the signature. Extremely rough estimates for these lifetimes are given in the Table. The lifetimes also appear in the appropriate versions of the Drake equation. This does raise a question about whether even a modified version of the Drake equation is appropriate for interstellar archaeology. For example, perhaps some factor proportional to the degree of difficulty should appear. More consideration needs to be given to this.

Annis [78] has raised the interesting possibility that his approach to looking for Kardashev Type III civilizations could be applied to study other astronomical distributions to look for non-natural outliers. Indeed, this embodies some of the spirit of recent discussions about blue stragglers by Livio [63] and Beech.

The presence of natural signatures that mimic interstellar archaeology signals is a significant problem. Both Dyson sphere searches and searches for artificially-driven blue stragglers are seriously compromised by natural signals. Conventional SETI is much better in this regard.

In short, interstellar archaeology has many problems. On the other hand the time may have come when interstellar archaeology including SETI should be considered seriously as a part of the web of science.

## ACKNOWLEDGEMENTS

The material presented here builds in part on foundations developed by G. Lemarchand and M. Beech. The author would like to thank J. Annis for guidance and help.

| | Examples of Interstellar Archaeology | | | | | | | |
|---|---|---|---|---|---|---|---|---|
| **Kar. num.** | **Interstellar archaeology type** | **Ref.** | **Reach (1000 ly)** | **$L_c$ (lifetime)** | **$L_c$ (kyrs)** | **Power Needs (W)** | **Mass Involved (kg)** | **Problems** |
| **0** | **SETI(radio)** | **[25]** | **to 0.25 now, 30 soon** | **civilization** | **5** | **$10^6$** | | **often needs intent** |
| **0** | **planetary atmospheres** | **[39]** | **O(0.1)** | **atmospheric perturbation** | **O(0.1)** | | **$\sim 10^{15}$** | **ambiguity** |
| **0** | **stellar salting** | **[43]** | **~30** | **λ isotope** | **$O(10^3)$** | | **$10^8$** | **natural signals** |
| **0** | **nuclear waste** | **[46]** | **~30** | **λ waste** | **$O(10^1)$** | | **$10^8$** | **ambiguity** |
| **I - II** | **spectral modulation** | | **60 (also ext. gal.)** | **civilization** | **5** | **$10^{26}$** | **$10^{24}$/yr** | **natural signals** |
| **II** | **Dyson sphere** | **[5]** | **to 1** | **civilization dyn. stab.** | **5** | **$4*10^{26}$** | **$10^{25}$** | **mimics** |
| **II** | **stellar engineering** | **[12]** | **20** | **~stellar lives** | **$10^6$** | **$4*10^{26}$** | **$10^{30}$** | **~~blue stragglers~~** |
| **II.5 - III** | **Fermi bubble** | | **$O(10^5)$** | **0.1 galaxy crossing** | **$10^4$** | **$10^{35}$** | **$10^{34}$** | **confusing signature** |
| **III** | **galactic Dyson sphere ensemble** | **[23]** | **$O(10^5)$** | **galaxy crossing** | **$10^5$** | **$10^{37}$** | **$10^{36}$** | **dark galaxies** |

Figure captions

Figure 1: Kardashev scale. Type I –power harnessed from a planet (NASA image), Type II, power from a star, perhaps via a Dyson sphere (based on schematic from Lemarchand), Type III, power from a galaxy. The Type III graph is from the Annis search. The scale could be extended to the visible universe or smaller systems.

Figure 2: IRAS 17446-4048, a typical carbon star. IRAS filter values: diamonds; DIRBE points: open circles; 2MASS: triangles; LRS (low): crosses; LRS (high): dots; fit to filter: dotted line; fit to LRS: solid line (scale: F(Jy) = y-axis/0.0046).

Figure 3: Closest approximation to a Dyson Sphere spectrum found in a recent search. The open squares are MSX points. The DIRBE points at low wavelength do not agree with the 2MASS values and are possibly from another source. (Same legend as Fig, 2, scale: F(Jy) = y axis/0.0168, IRAS 20369+5131).

Fig. 4: Galactic Aitoff plot for the sixteen sources from a recent Dyson sphere search. Three lowest least squares sources – squares, others - circles (one is shadowed). Sources in a selected 2240 source sample - dots. The SETI Arecibo region lies between the 38º and -2º bounds. The galactic plane is clear in the distribution but not the galactic bulge.

Figures:

Figure 1

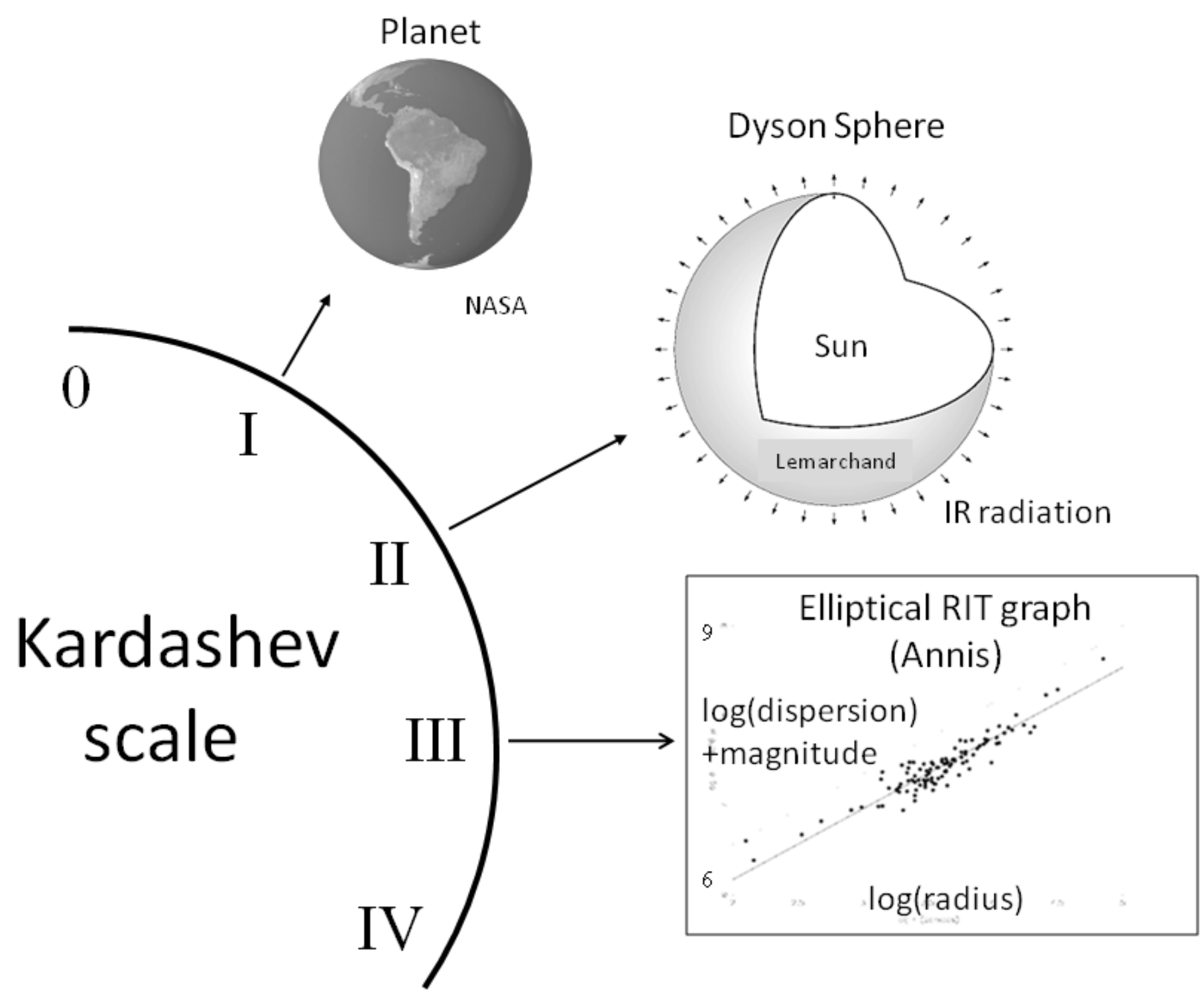

Figure 2

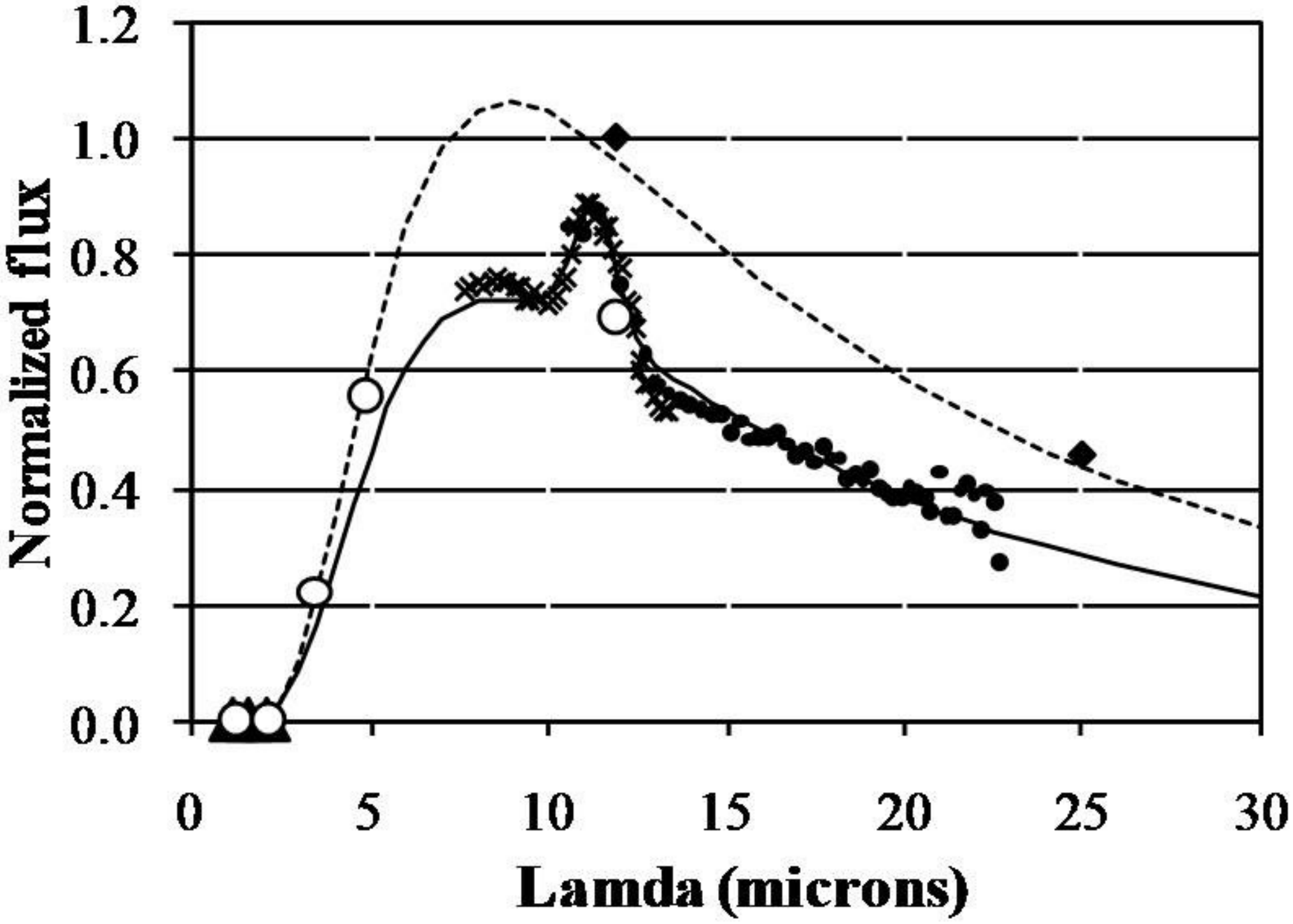
1.2
1.0
0.8
0.6
0.4
0.2
0.0
Normalized flux
0
5
10
15
20
25
30
Lamda (microns)

Figure 3

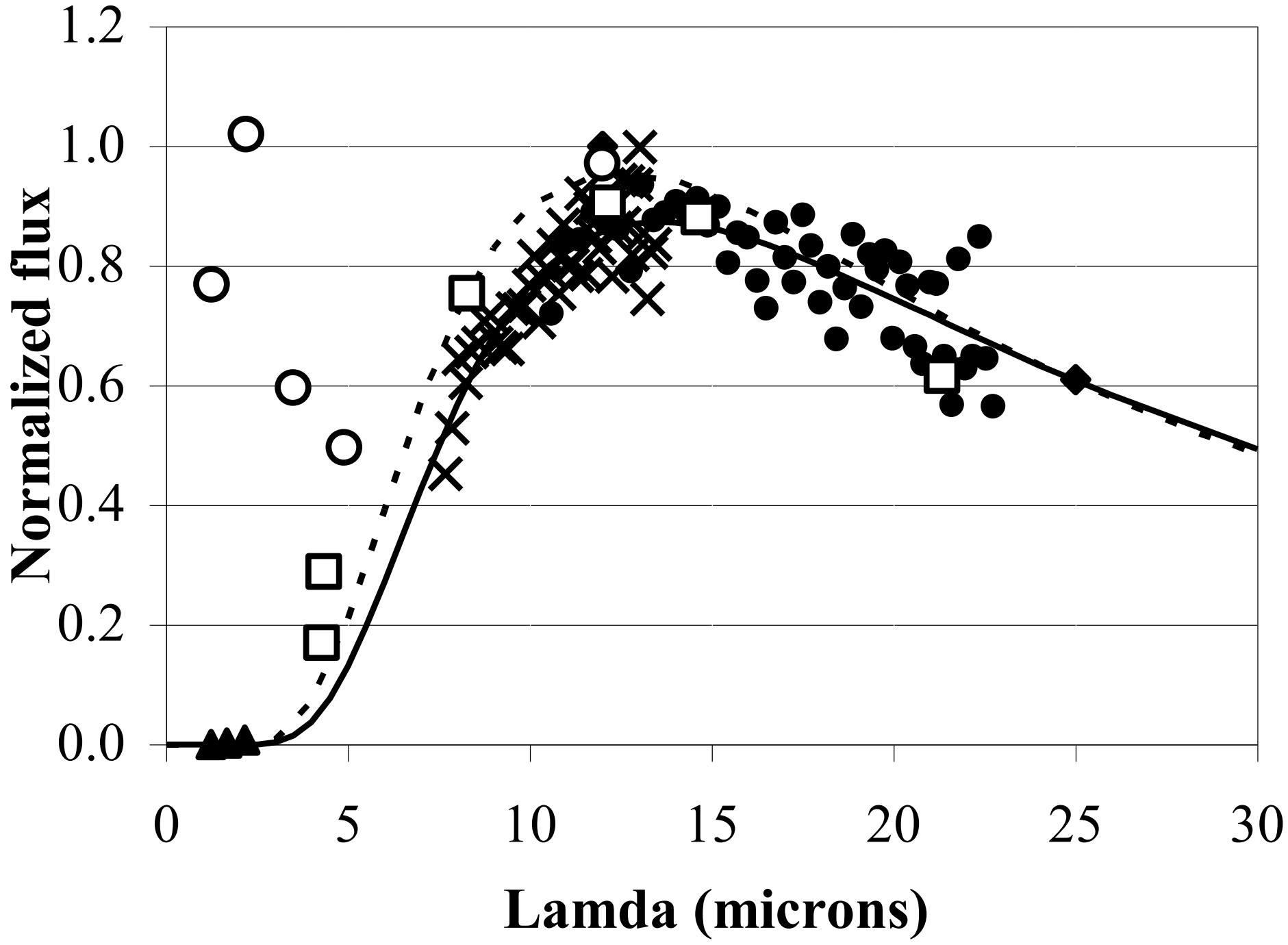

1.2
1.0
0.8
0.6
0.4
0.2
0.0
Normalized flux
0
5
10
15
20
25
30
Lamda (microns)

Figure 4

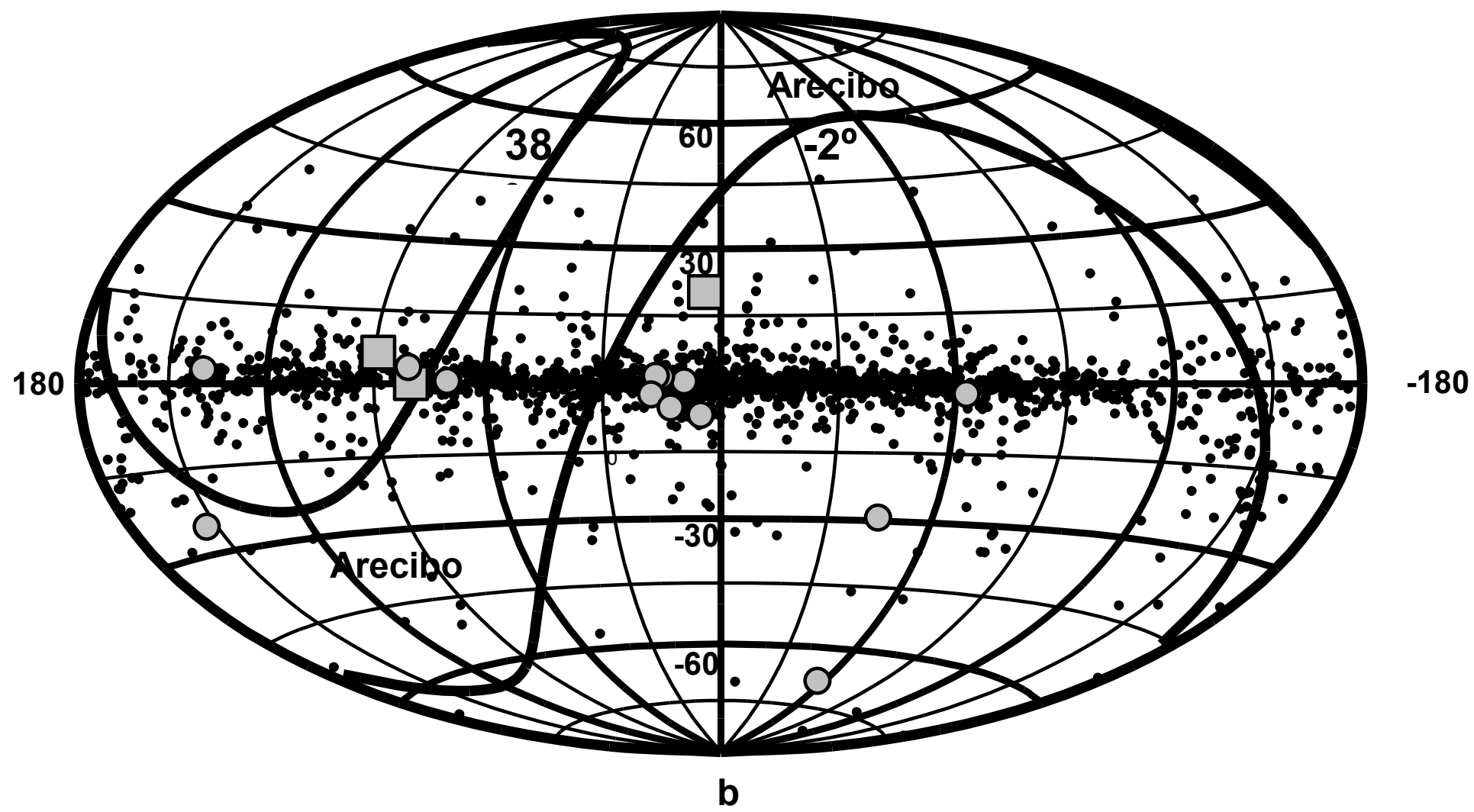